\documentclass{aa}  
\usepackage[varg]{txfonts}      

\usepackage{natbib,twoopt}
\usepackage[breaklinks=true]{hyperref} 
\bibpunct{(}{)}{;}{a}{}{,}             
\makeatletter
  \newcommandtwoopt{\citeads}[3][][]{\href{http://adsabs.harvard.edu/abs/#3}%
    {\def\hyper@linkstart##1##2{}%
     \let\hyper@linkend\@empty\citealp[#1][#2]{#3}}}
  \newcommandtwoopt{\citepads}[3][][]{\href{http://adsabs.harvard.edu/abs/#3}%
    {\def\hyper@linkstart##1##2{}%
     \let\hyper@linkend\@empty\citep[#1][#2]{#3}}}
  \newcommandtwoopt{\citetads}[3][][]{\href{http://adsabs.harvard.edu/abs/#3}%
    {\def\hyper@linkstart##1##2{}%
     \let\hyper@linkend\@empty\citet[#1][#2]{#3}}}
  \newcommandtwoopt{\citeyearads}[3][][]%
    {\href{http://adsabs.harvard.edu/abs/#3}
    {\def\hyper@linkstart##1##2{}%
     \let\hyper@linkend\@empty\citeyear[#1][#2]{#3}}}
\makeatother

\usepackage{graphicx}
\usepackage{dcolumn}
\usepackage[T1]{fontenc}
\usepackage[utf8]{inputenc}
\usepackage{wrapfig} 
\usepackage{amsmath} 
\usepackage{cancel} 
\usepackage{amssymb} 
    \usepackage{bm}  
\usepackage{multirow} 

\usepackage{varwidth}
\usepackage[svgnames]{xcolor} 
\usepackage{subcaption}
\usepackage{tabularx}
\usepackage{array} 
\newcolumntype{L}[1]{>{\raggedright\let\newline\\\arraybackslash\hspace{0pt}}m{#1}}
\newcolumntype{C}[1]{>{\centering\let\newline\\\arraybackslash\hspace{0pt}}m{#1}}
\newcolumntype{R}[1]{>{\raggedleft\let\newline\\\arraybackslash\hspace{0pt}}m{#1}}
\usepackage{siunitx}
    \DeclareSIUnit\erg{erg}
\begin{document}

\title{The effect of magnetic field on the damping of slow waves in the solar corona}

\author{T.~J.~Duckenfield\inst{1}
        \and
        D.~Y.~Kolotkov\inst{1,} \inst{2}
        \and
        V.~M.~Nakariakov\inst{1,}\inst{3}
        }

\institute{Centre for Fusion, Space and Astrophysics, Department of Physics, University of Warwick, CV4 7AL, UK \\
           \email{T.Duckenfield@warwick.ac.uk} 
           \and
           Institute of Solar-Terrestrial Physics SB Russian Academy of Sciences, Irkutsk 664033, Russia
           \and 
           St Petersburg Branch, Special Astrophysical Observatory, Russian Academy of Sciences, St Petersburg, 196140, Russia
           }
           
\date{Received 2 November 2020 / Accepted 20 November 2020} 

\abstract{
    Slow magnetoacoustic waves are routinely observed in astrophysical plasma systems such as the solar corona, and are usually seen to damp rapidly. As a slow wave propagates through a plasma, it modifies the equilibrium quantities of density, temperature, and magnetic field. 
    In the corona and other plasma systems, the thermal equilibrium is comprised of a balance between continuous heating and cooling processes, the magnitudes of which vary with density, temperature and magnetic field. 
    Thus the wave may induce a misbalance between these competing processes. 
    Its back reaction on the wave has been shown to lead to dispersion, and amplification or damping, of the wave. }
    { 
    This effect of heating/cooling misbalance has previously been studied in the infinite magnetic field approximation, in a plasma whose thermal equilibrium comprises of optically thin radiative losses and field-aligned thermal conduction, balanced by an (unspecified) heating process.
    In this work we extend this analysis by considering a non-zero $\beta$ plasma.
    The importance of the effect of magnetic field in the rapid damping of slow waves in the solar corona is evaluated, and compared to the effects of thermal conduction.} 
    { 
    A linear perturbation under the thin flux tube approximation is considered, and a dispersion relation  describing the slow magnetoacoustic modes is found.
    The dispersion relation's limits of strong non-adiabaticity and weak non-adiabaticity are studied.
    The characteristic timescales are calculated for plasma systems with a range of typical coronal densities, temperatures and magnetic field strengths.
    }
    { 
    The number of timescales characterising the effect of misbalance is found to remain at two, as with the infinite magnetic field case. 
    In the non-zero $\beta$ case, these two timescales correspond to the partial derivatives of the combined heating/cooling function with respect to constant \textit{gas} pressure and with respect to constant \textit{magnetic} pressure. 
    The predicted damping times of slow waves from thermal misbalance in the solar corona are found to be of the order of 10--100 minutes, coinciding with the wave periods and damping times observed. 
    Moreover the slow wave damping by thermal misbalance is found to be comparable to the damping by field-aligned thermal conduction. 
    The change in damping with plasma-$\beta$ is complex and depends on the coronal heating function's dependence upon the magnetic field in particular.
    Nonetheless we show that in the infinite field limit, the wave dynamics is insensitive to the dependence of the heating function on the magnetic field, and this approximation is found to be valid in the corona so long as the magnetic field strength is greater than approximately \SI{10}{G} for quiescent loops and plumes, and \SI{100}{G} for hot and dense loops.
     }
    { 
    Thermal misbalance may damp slow magnetoacoustic waves rapidly in much of the corona, and its inclusion in our understanding of slow mode damping may resolve discrepancies between observations and theory relying on compressive viscosity and thermal conduction alone. } 
    
   \maketitle          
%
\section{Introduction}
\label{sec:intro}
Slow magnetoacoustic waves are a common feature of many plasma systems, 
and the study of their properties allows one to probe the local plasma conditions which otherwise may be difficult to measure. 
Often these plasma systems are maintained at thermal equilibrium by a delicate balance between continuous heating and cooling mechanisms - one example being the solar corona, which is cooled by radiative losses and heated by some as-yet undetermined heating process(es). 
The effect of these heating and cooling mechanisms vary with the plasma parameters. 
As a slow wave propagates through the plasma, the wave perturbs both the plasma's mechanical and thermal equilibria, through modifications in the local density, temperature and magnetic field strength. 
Thus if the plasma is steadily being heated and cooled at thermal equilibrium, the wave induces a misbalance between these competing processes. 
This leads to the transfer of energy between the wave and the plasma referred to as a \emph{heating/cooling misbalance}.
 
Previous studies of the effects of this wave-induced thermal misbalance under the infinite magnetic field approximation have shown that the plasma may act as a dissipative or active medium, damping the wave or growing its amplitude \citep{Nakariakov2000, Kumar2016}. 
The presence of characteristic times associated with the thermal misbalance may also cause dispersion, such that any broadband pulse is dispersed by the medium into a quasi-periodic slow wave train \citep{Zavershinskii2019}.
Such observable effects upon the wave by the heating/cooling misbalance are related to the properties of the heating and cooling processes themselves, specifically their derivatives with respect to the thermodynamic parameters of the plasma: density $\rho$, temperature $T$ and potentially magnetic field strength $B$.
In \citet{Kolotkov2019} the damping by thermal misbalance of hot coronal loops observed by SUMER was considered, and it was found that in the regime of enhanced damping, the theoretically obtained damping rates coincide with those seen in observations.  
Recently, \citet{Kolotkov2020_constrainH} demonstrated the potential for constraining the unknown coronal heating function, using observations of the solar corona such as the observed rapid damping of slow modes, and coronal slow waves' thermal instability and acoustic overstability.


These previous works rest on the assumption of infinitely strong magnetic field and consider perturbations to the local plasma density and temperature, following from the seminal work of \citet{Field1965} analysing perturbations to an infinite homogeneous plasma. 
Yet to fully understand the effects of the heating/cooling misbalance, the influence of non-zero $\beta$ must be studied, since any non-zero magnetic field fluctuations will interact with the density, temperature and velocity perturbations and therefore affect the wave evolution and propagation speed \citep{Afanasyev2015,Nakariakov2017}. 
Some magnetic field measurements of coronal structures have found magnetic field strengths can drop to \SI{10}{G} and below \citep[e.g. a value of \SI{4}{G} reported in][]{Lin2004}, implying that the magnetic pressure may not totally dominate over gas pressure everywhere. 
Moreover, the source of energy for the solar coronal heating is known to be the magnetic field, and so it is natural to allow a dependence of the heating/cooling misbalance upon the magnetic field strength. 
It is therefore important to investigate the role magnetic effects have upon the dispersion and damping by heating/cooling misbalance.

The rapid damping of slow magnetoacoustic modes observed in the solar corona is well documented, yet ambiguity remains regarding its origin \citep[e.g.][]{DeMoortel2009}. 
Thermal conduction and compressive viscosity are invoked as damping mechanisms, however we contend that the effect of wave-induced heating/cooling misbalance can be of equal importance.
The inclusion of thermal misbalance as an additional damping mechanism could resolve discrepancies seen in the frequency dependencies of observed slow mode damping, such as for the standing slow modes of hot loops reported in \citet{Mariska2006}, the propagating slow modes in coronal holes detailed in \citet{Gupta2014}, and the propagating slow modes in the warm corona analysed in \citet{Prasad2014}. 
The phase shifts between density and temperature measured in e.g. \citet{Prasad2018} disagree with those predicted from theory \citep{Owen2009}, and similarly the growth in polytropic index (estimated via phase shift) with temperature observed in \citet{VanDoorsselaere2011, Prasad2019} are also a mystery.
The series of papers culminating in \citet{Wang2019_pt3} try to rectify these and similar discrepancies between 1D slow mode damping theory and observations through anomalous thermal conduction and viscosity coefficients -- the inclusion of thermal misbalance provides an alternative, perhaps more physically motivated, explanation.

In this work we extend the results of \citet{Kolotkov2019} to investigate the effects of thermal misbalance in non-zero $\beta$ plasma upon a slow wave using the thin flux tube approximation. 
A non-adiabatic linear dispersion relation is derived, and its limits of weak and strong non-adiabaticity are explored in Section~\ref{sec:dispersion_relation}.
Estimates of the damping time of slow waves in the solar corona by thermal misbalance, its sensitivity to the dependence of the heating/cooling function on $B$, and comparisons with other dissipation mechanisms are the topics of Section~\ref{sec:slow_mode_damping}. 
Discussion and concluding remarks are made in Section~\ref{sec:discussion}. 

\section{Dispersion relation}
\label{sec:dispersion_relation}
\subsection{Derivation}
\label{subsec:derivation}
In this work we use the first order thin flux tube approximation, which formally corresponds to the first order of the Taylor expansion of the MHD variables with respect to the radial coordinate, derived by \citet{Roberts1978} and \citet{Zhugzhda1996}. 
The governing equations are the same as for \citet{Nakariakov2017}, neglecting the viscous dissipation in the momentum equation and slightly adjusting the definition of the thermal heating/cooling function $Q$ to have the units of W~kg\textsuperscript{-1} \citep[matching the definition in][]{Field1965,Kolotkov2019}: 
\begin{align} 
&\frac{dp}{dt}  -  \frac{\gamma p}{\rho}\frac{d\rho}{d t} = \left( \gamma - 1 \right) \left( \kappa_\parallel \frac{\partial^2 T}{\partial z^2} - \rho Q(\rho,T,B) \right),  \label{eqn:misbalance_gov_energy} \\
&\rho \frac{d u}{d t} + \frac{\partial p}{\partial z}= 0,\label{eqn:misbalance_gov_mmtm} \\
&p + \frac{B^2}{2 \mu_0} = p_\text{total}^\text{ext},  \label{eqn:misbalance_gov_radial} \\
&\frac{\partial B}{\partial t} + u\frac{\partial B}{\partial z} + 2 B v_r = 0 \label{eqn:misbalance_gov_induction} \\
&\frac{\partial \rho}{\partial t} + 2 \rho v_r +\frac{\partial}{\partial z}(\rho u) = 0, \label{eqn:misbalance_gov_continuity} \\
& p = \frac{k_\mathrm{B}}{m}\rho T  \label{eqn:misbalance_gov_state}
\end{align}
As usual, $p$ is the plasma pressure, $\rho$ is the plasma density, $T$ is the temperature, $k_\mathrm{B}$ is the Boltzmann constant, $\mu_0$ is the magnetic permeability of free space, $m$ is the mean particle mass, and $\gamma$ is the polytropic index. 
Also $u$ is the wave-induced flow speed along the tube (i.e. in $z$ direction), $B$ is the component of magnetic field strength along the tube, $p_\text{total}^\text{ext}$ is the total external pressure and $v_r$ is the radial derivative of the radial component of plasma velocity. All of these quantities are measured at the axis of the (infinitesimally thin) flux tube. 
The right hand side of the energy equation (\ref{eqn:misbalance_gov_energy}) represent thermodynamic processes ongoing inside the plasma. 
The first term is the (field-aligned) thermal conduction, for which we use the standard estimation of coefficient $\kappa_\parallel \approx 10^{-11} T^{5/2}$~Wm\textsuperscript{-1}K\textsuperscript{-1}. 
The second term is the combination of some unspecified heating $\mathcal{H}(\rho,T,B)$ and optically thin radiative cooling $\mathcal{L}(\rho,T)$, combined in the net heat/loss function $Q(\rho,T,B) = \mathcal{L} - \mathcal{H}$. 
We note $Q$ depends on $B$ only if the heating term $\mathcal{H}$ is a function of $B$, since the radiative losses $\mathcal{L}$ are known to be independent of $B$. 

Thus, in addition to the perturbation of the mechanical equilibrium provided by the force balance, in this work we consider a wave-induced perturbation of the thermal equilibrium of the corona. 
It is important to remark explicitly that, following from the previous works on thermal misbalance, we allow \emph{both} the heating and cooling functions to be perturbed. 
This is is contrast to several previous works in which the heating term is held constant, which is to say remains unperturbed by the wave, such as \citet{Claes2019,Kaneko2017,DeMoortelHood2003_pt1} do when setting up their simulations.

We consider linear perturbations of a mechanical equilibrium, characterised by the constant quantities denoted $p_0$, $\rho_0$, $B_0$, $T_0$, and $p_\text{total}^\text{ext}$, and without steady flows.
In addition we consider $Q_0=0$ in the equilibrium, motivated by the continued existence of the corona. 
The parallel thermal conduction does not contribute to this equilibrium because the plasma temperature is uniform. 
Let the perturbations of the equilibrium quantities be small,
\begin{align*}
    &p = p_0 + p_1, \ \rho = \rho_0 + \rho_1, \ T = T_0 + T_1, \\
    &B = B_0 + B_1, \ v_r = v_1, \ u = u_1,
\end{align*}
where the subscript 1 denotes small perturbations. 
In the following, exceptions are made for $v_r,u$ since these are small quantities about zero anyway and so we leave their subscripts alone. 
We substitute these quantities into Equations~(\ref{eqn:misbalance_gov_energy}) - (\ref{eqn:misbalance_gov_state}) and keep only the linear terms of the small quantities to find:
\begin{align} 
    \begin{split}
        &\frac{\partial}{\partial t}p_1  -  C_\mathrm{S}^2\frac{\partial}{\partial t}\rho_1 = \\
        & \qquad \left( \gamma - 1 \right) \left( \kappa_\parallel \frac{\partial^2}{\partial z^2}T_1 - \rho_0\left[ Q_\rho \rho_1 + Q_T T_1 + Q_B B_1 \right] \right), 
    \end{split}\label{eqn:misbalance_linear_energy} \\
    \begin{split} 
        &\rho_0 \frac{\partial}{\partial t} u + \frac{\partial}{\partial z} p_1 = 0,
    \end{split}\label{eqn:misbalance_linear_mmtm} \\
    \begin{split}
        &p_1 + \frac{B_0 B_1}{\mu_0} = 0,
    \end{split} \label{eqn:misbalance_linear_radial} \\
   \begin{split} 
        &\frac{\partial}{\partial t} B_1 + 2 B_0 v_r = 0,
    \end{split} \label{eqn:misbalance_linear_induction} \\
    \begin{split}
        &\frac{\partial}{\partial t} \rho_1 + 2 \rho_0 v_r + \rho_0 \frac{\partial}{\partial z}u = 0,
    \end{split} \label{eqn:misbalance_linear_continuity} \\
    \begin{split}
        & p_1 - \frac{k_\mathrm{B}}{m}\left( \rho_0 T_1 + T_0 \rho_1 \right) = 0. 
    \end{split}\label{eqn:misbalance_linear_state} 
\end{align}
The parameter $C_\mathrm{S}^2 = \gamma p_0/\rho_0$ is the sound speed at equilibrium, and $Q_T, Q_\rho, Q_B$ are the partial derivatives of the combined heating/cooling function $Q$ ($Q_x = \partial Q/\partial x$), evaluated at the equilibrium. 

By using these equations, several assumptions have been made which are worth mentioning. 
Firstly, since in a slow wave in a low-beta plasma, any change in the external pressure $p_\text{total}^\text{ext}$ to the flux tube is neglected \citep[e.g. see][]{EdwinRoberts1983}, we concentrate on waves propagating inside the flux tube, taking that the slow waves are always in the trapped regime. 
Secondly, the obliqueness of the wavefronts are accounted for through the use of $v_r$ -- this is valid when the wavelength of the perturbations (parallel to the field) is much longer than the transverse spatial scale, determined by the width of the waveguiding plasma non-uniformity. This is the applicability condition of the thin flux tube approximation, and is a key difference to the plane acoustic wave case used elsewhere. 
Finally it should be noted that, for all non-adiabatic processes in general (whose assorted characteristic timescales we call $\tau_i$), the assumption $\gamma = C_P/C_\mathrm{V}$ is only valid when $\omega \gg \tau_i^{-1}$ , i.e. when the wave is of sufficiently high frequency that it is adiabatic or weakly non-adiabatic \citep[for example see the discussion in][]{VanDoorsselaere2011}. 
In general non-adiabatic scenarios \citep{Zavershinskii2019} or non-zero $\beta$ plasmas \citep{Nistico2017}, the ratio of specific heats can vary. 
In our estimations, we use $\gamma = 5/3$ without loss of generality. 

Since we have no flows and we have uniformity in the $z$ direction, we take a Fourier transform by making the ansatz of plane waves, that is to say we assume a harmonic dependence upon the time and spatial coordinates for all perturbed variables $\propto \exp{(-i \omega t + i k z)}$ where $\omega$ is the frequency and $k$ is the parallel wavenumber. 
The resulting linearised set of equations yield the following dispersion relation:
\begin{equation}
    \omega^3 + A(k)\omega^2 + B(k)\omega + C(k) = 0 \label{eqn:dispersion_relation_1}
\end{equation}
where the coefficients are 
\begin{align*}
    A &= i \frac{C_\mathrm{T}^2}{C_\mathrm{S}^2}\Bigg\{ \frac{Q_T}{C_\mathrm{V}} + \frac{\kappa_\parallel }{\rho_0 C_\mathrm{V}}k^2 + \frac{(\gamma - 1)}{C_\mathrm{A}^2}\Big[ T_0 Q_T - B_0 Q_B -\rho_0 Q_\rho \\
    &\quad + \frac{T_0}{\rho_0} \kappa_\parallel k^2 \Big] \Bigg\}, \\
    B &= - C_\mathrm{T}^2 k^2, \\
    C &= - i(\gamma -1)\frac{C_\mathrm{T}^2}{C_\mathrm{S}^2}\Big(T_0 Q_T - \rho_0 Q_\rho +  \frac{T_0}{\rho_0}\kappa_\parallel k^2 \Big)k^2 =0.   
\end{align*}
The term $C_\mathrm{A}$ is the standard Alfv\'en speed defined through $C_\mathrm{A}^2 =  {B_0^2}/{\mu_0 \rho_0}$, and the term $C_\mathrm{T}$ is the tube speed defined through $C_\mathrm{T}^{-2} = C_\mathrm{S}^{-2} + C_\mathrm{A}^{-2}$.  
Equation~(\ref{eqn:dispersion_relation_1}) is cubic in $\omega$ yet quartic in $k$, that is to say asymmetric with respect to space and time.
This dispersion relation describes two oppositely-directed propagating slow waves and an entropy mode, made into a thermal mode by the non-adiabatic effects \citep[e.g.][]{DeMoortelHood2003_pt1}.
The tube speed appears in the coefficient of the ${\omega}$ term, thus in the adiabatic limit  
the equation reduces to the wave equation with $C_\mathrm{T}^2$ as the speed, as expected for so-called tube waves. 
This expression also agrees exactly with the infinite magnetic field case \citep[Eq.~7,][]{Kolotkov2019} in the limit $B\rightarrow \infty$.

Regarding the thermal conduction terms, we see the term $(C_\mathrm{T}^2/C_\mathrm{S})^2 \kappa_\parallel k^2/\rho_0 C_\mathrm{V}$ in the $\omega^2$ coefficient $A$, which is proportional to the term in the infinite magnetic field case \citep[e.g. Eq.~8 in][]{Kolotkov2019} but modified by the ratio of tube to sound speed squared. 
Thus there is a non-zero $\beta$ modification to the effect by thermal conduction on the waves, which is qualitatively consistent with the result in \citet{Afanasyev2015}. 

It is convenient to re-express the non-adiabatic terms using characteristic timescales, which for the thermal misbalance terms are fully determined by the equilibrium parameters and partial derivatives $Q_\rho, Q_B, Q_T$. 
Note that these timescales are \emph{not} determined by the heating and cooling processes separately. 
Rather, these characteristic timescales are determined by how quickly the perturbation returns to, or destroys, the equilibrium. 

In the case with infinite magnetic field, the characteristic timescales were written in terms of $Q_{T[p]}$ and $Q_{T[\rho]} = Q_T$, where $Q_{T[p]}$ means the partial derivative taken with respect to temperature assuming constant gas pressure \citep{Zavershinskii2019}. 
The introduction of a finite magnetic field means there is a separate, magnetic pressure term $B^2/2\mu_0$. 
Thus we consider separately the derivative with respect to constant gas pressure $Q_{T[\text{gas }p]}$, and with respect to constant magnetic pressure $Q_{T[\text{mag }p]}$. 
To write the additional terms in the derivatives of $Q$ by the magnetic field, the relevant equations are the radial pressure balance (Eq.~\ref{eqn:misbalance_linear_radial}) and the ideal gas law (Eq.~\ref{eqn:misbalance_linear_state}), finding:
\begin{equation*}
\left.
\begin{aligned}
    & \frac{\partial \rho}{\partial T} = -\frac{m}{k_\mathrm{B}}\frac{p_0}{T_0^2} = -\frac{\rho_0}{T_0}, \\ 
    & \frac{\partial p}{\partial T} = \frac{k_\mathrm{B}}{m}\rho_0, \\ 
    & \frac{\partial B}{\partial T} = -\frac{\mu_0}{B_0}\frac{\partial p}{\partial T} = -\frac{\beta}{2}\frac{B_0}{T_0}
\end{aligned}
\quad \right\} \implies 
\begin{aligned}
    &Q_{T [\text{gas }p]}  \, = Q_T - \frac{\rho_0}{T_0}Q_\rho \\
    &Q_{T [\text{mag }p]} \, = Q_T - \frac{\beta}{2}\frac{B_0}{T_0}Q_B.
\end{aligned}  \label{eqn:partials}
\end{equation*}
Gathering terms in the dispersion relation (Eq.~\ref{eqn:dispersion_relation_1}) we can define two characteristic timescales,
\begin{align}
    \tau_1 &= \frac{C_P}{Q_{T[\text{gas } p]}} = \frac{\gamma C_\mathrm{V}}{Q_{T[\text{gas } p]}}, & \tau_2 &= \frac{C_\mathrm{V}}{Q_{T[\text{mag } p]}}. 
    \label{eqn:misbalance_timescales}
\end{align}
It is striking that despite there being three different dependencies in $Q(\rho,T,B)$, the effects of heating/cooling misbalance can be expressed in terms of these two timescales evaluated at constant gas and magnetic pressures.
Comparing these timescales with the infinite magnetic field case \citep{Kolotkov2019,Zavershinskii2019} we see that $\tau_1$ is identical, whilst $\tau_2$ is different only by a magnetic correction term in $Q_{T[\text{mag } p]}$, which goes to $Q_T$ as the plasma-$\beta$ goes to zero.
We also use the characteristic timescale for thermal conduction (in the infinite magnetic field limit) - for wavelength $\lambda = 2\pi/k$ -  as given in \citet{Kolotkov2019}, namely $\tau_\mathrm{cond}(k) = \rho_0 C_\mathrm{V} \lambda^2 /\kappa_\parallel$.
Pulling these definitions together, we write the dispersion relation as
\begin{equation}
    \begin{split} &
    {\omega^3} + i\frac{2}{2+\gamma \beta}\left\{ \frac{4\pi^2}{\tau_\text{cond}(k)}\left( 1 + \frac{\beta}{2} \right) + \frac{1}{\tau_2} + \frac{\gamma\beta }{2}\frac{1}{\tau_1} \right\} {\omega^2} \\ &
    - C_\mathrm{T}^2 k^2 {\omega} - i C_\mathrm{T}^2\left\{\frac{1}{\gamma}\frac{4\pi^2}{\tau_\mathrm{cond}(k)} + \frac{1}{\tau_1} \right\}k^2 =0.     
    \end{split}
    \label{eqn:dispersion_relation_final}
\end{equation}
It may be seen that Equation~(\ref{eqn:dispersion_relation_final}) is affected by the magnetic field in several ways: the phase speed $C_\mathrm{S} \rightarrow C_\mathrm{T}$, the terms with plasma-$\beta$, and also implicitly through the timescale $\tau_2$ (i.e. via the product $\beta Q_B/2$). 
Interestingly, only the last of these is affected by the dependence of $Q$ upon magnetic field $B$. 
This means that \emph{even if the heating model is independent of magnetic field, the properties of the wave are still affected by the magnetic field strength.}
The reverse is also true, if $\beta$ goes to zero the wave dynamics is not affected by the magnetic field even if the heating function has some dependence on it.

The non-zero $\beta$ effects on the real part of $\omega$ (and hence phase speed) are well known, and it has been demonstrated they may be important for slow waves in some wave guides such as hot flaring loops \citep{Afanasyev2015}, as well as the determination of cut-off frequency in the solar atmosphere. 
By accounting for the obliqueness of the waves, the wave speed is made to depend on the absolute value of the magnetic field via $C_\mathrm{T}$, which is sub-sonic and sub-Alfv\'enic. 

\subsection{Limit of weak non-adiabaticity}
\label{subsec:weak_limit}
Similar to the previous works on damping of magnetoacoustic waves by thermal conduction
, the upper and lower limits of non-adiabaticity are now derived. 
We begin with the limit of weak non-adiabaticity, in which the wave is only mildly affected by the transfer of energy with the active medium. 
In this limit $\omega \gg 1/\tau_{1,2,\mathrm{cond}}$, thus we rearrange the dispersion relation~(\ref{eqn:dispersion_relation_final}) assuming $\omega \ne 0$ into
\begin{equation}
    \begin{split}
        \omega^2 = C_\mathrm{T}^2 k^2  \Bigg\{1 & - i\Bigg[ \frac{1}{C_\mathrm{T}^2}\frac{\omega^2}{k^2}\left(\frac{1}{\omega \tau_2} + \frac{\gamma\beta}{2}\frac{1}{\omega \tau_1} \right)\frac{C_\mathrm{T}^2}{C_\mathrm{S}^2} - \frac{1}{\omega \tau_1} \\
            & + \frac{4\pi^2}{\omega \tau_\mathrm{cond}(k)}\left( \frac{1}{C_\mathrm{T}^2}\frac{\omega^2}{k^2}\left[1 + \frac{\beta}{2} \right]\frac{C_\mathrm{T}^2}{C_\mathrm{S}^2}  - \frac{1}{\gamma} \right) \Bigg] \Bigg\},
    \end{split}
\end{equation}
\citep[consistent with Eq.~(21)][which deals with the same limiting case]{Nakariakov2017}. 
Taking the limit of $1/\omega\tau_\mathrm{cond}$ and $1/\omega\tau_{1,2}$ as small parameters, the Taylor expansion of the dispersion relation reduces to
\begin{equation}
    \omega^2 \approx C_\mathrm{T}^2 k^2  \Bigg\{1 - i\omega^{-1} \frac{2}{2+\gamma \beta}\Bigg[\left(\frac{\gamma-1}{\gamma}\right)\frac{4\pi^2}{\tau_\mathrm{cond}} + \frac{1}{\tau_2} - \frac{1}{\tau_1} \Bigg] \Bigg\}.
    \label{eqn:misbalance_ready4weak}
\end{equation}
\noindent To evaluate the $\omega^{-1}$ in the imaginary component of Equation~(\ref{eqn:misbalance_ready4weak}), perturbation theory is used. 
In the zeroth order $\omega \approx C_\mathrm{T} k$, and using this yields the following solution to the weakly non-adiabatic dispersion relation: 
\begin{align}
&\omega_\mathrm{R} \approx C_\mathrm{T} k, \label{eqn:misbalance_weak_Re} \\
&\omega_\mathrm{I} \approx - \frac{1}{2}\left(\frac{2}{2 + \gamma \beta }\right)\left[\frac{\gamma - 1}{\gamma}\frac{4\pi^2}{\tau_\mathrm{cond}}+\frac{1}{\tau_2}-\frac{1}{\tau_1}\right]. \label{eqn:misbalance_weak_Im}
\end{align}
The phase speed of the weakly non-adiabatic wave is the tube speed $C_\mathrm{T}$ as expected, so the phase speed is reduced as $\beta$ increases.
In the limit of $\beta \rightarrow 0$ this equation coincides with the results in \citet[][]{Kolotkov2019}. 

From Equation~(\ref{eqn:misbalance_weak_Im}) we are motivated to form the single, combined timescale $\mathcal{T}_\mathrm{M}(\tau_1, \tau_2)$ which can be referred to as a characteristic damping time of the heating/cooling misbalance in the weakly non-adiabatic regime:
\begin{equation}
    \begin{aligned}
    \frac{1}{\mathcal{T}_\mathrm{M}} &= \frac{C_\mathrm{T}^2}{C_\mathrm{S}^2}\left(\frac{1}{\tau_2} - \frac{1}{\tau_1} \right), \\
    &= \left(\frac{2}{2 + \gamma \beta }\right)\left(\frac{1}{\tau_2} - \frac{1}{\tau_1}\right), \\
    &= \frac{2}{2 + \gamma \beta }\left\{\frac{\gamma-1}{\gamma}\frac{Q_T}{C_\mathrm{V}} + \frac{1}{\gamma}\frac{\rho_0}{T_0}\frac{Q_\rho}{C_\mathrm{V}}  
    -\frac{\beta}{2}\frac{B_0}{T_0}\frac{Q_B}{C_\mathrm{V}}\right\}.
    \end{aligned} \label{eqn:tau_M} 
\end{equation}
From a physical standpoint, this equation implies that in the considered limit of weak non-adiabaticity, the perturbations of $T$, $\rho$, $B$ all affect the heat/loss function $Q$ independently (i.e. the effects of $Q_T$, $Q_\rho$ and $Q_B$ are additive in Eqs.~\ref{eqn:misbalance_weak_Im} and \ref{eqn:tau_M}).
In the absence of thermal conduction, the $e$-folding time of the slow wave amplitude by thermal misbalance is accordingly $\mathcal{T}_\mathrm{damp} = 2\mathcal{T}_\mathrm{M}$, consistent with the relationship between damping time and characteristic times defined in \citet{DeMoortelHood2003_pt1}. 
In a similar fashion, we form a timescale $\mathcal{T}_\mathrm{condB}(k,\beta)$ characteristic of the wave damping time by thermal conduction in the limit of weak non-adiabaticity, related to $\tau_\mathrm{cond}(k)$ by
\begin{equation} \label{eqn:tau_condB}
    \mathcal{T}_\mathrm{condB} = \left(1 + \frac{\gamma \beta }{2}\right)\frac{\gamma}{\gamma - 1}\frac{\tau_\mathrm{cond}(k)}{4\pi^2}, 
\end{equation}
\begin{equation}
\implies \ \omega_\mathrm{I} \approx -\frac{1}{2}\left(\frac{1}{\mathcal{T}_\mathrm{condB}} + \frac{1}{\mathcal{T}_\mathrm{M}}\right). \label{eqn:tau_M_adds}
\end{equation}
If the limit of weak non-adiabaticity applies, there is no thermal conduction, and $\mathcal{T}_\mathrm{M}>0$, then we may \emph{interpret $\mathcal{T}_\mathrm{M}$ as the damping time over which the thermal misbalance is attenuating the slow wave}. 
Similarly, if the limit of weak non-adiabaticity applies and $\mathcal{T}_\mathrm{M}<0$, energy is supplied from the medium into the wave, amplifying the slow wave over the characteristic timescale $\mathcal{T}_\mathrm{M}$. 
Regarding the effect of non-zero $\beta$, the phase speed is reduced as $\beta$ increases through $C_\mathrm{T}$, and the effect on the damping depends on the sign of $Q_B$ -- note if there is no magnetic dependence ($Q_B=0$), then the damping effect is always lessened with increasing $\beta$.

The effect of thermal conduction is always to damp ($\mathcal{T}_\mathrm{condB}$ is strictly positive).
From Equation~(\ref{eqn:tau_condB}) it is clear that \emph{in the limit of weak non-adiabaticity, the damping effect of thermal conduction is reduced as the plasma-$\beta$ increases.}
In the weak non-adiabatic limit, the effect of thermal conduction and the effect of thermal misbalance upon the wave increment $\omega_\mathrm{I}$ are additive, shown in Equation~(\ref{eqn:tau_M_adds}). 

In the presence of both thermal conduction and thermal misbalance, whilst remaining in the weakly non-adiabatic limit, the damping time of the wave is $\mathcal{T}_\mathrm{damp} = 2/(\mathcal{T}_\mathrm{condB}^{-1} + \mathcal{T}_\mathrm{M}^{-1})$, equivalently

\begin{equation} \label{eqn:weak_damping_time}
    \mathcal{T}_\mathrm{damp} = \frac{\displaystyle2+\gamma \beta}{\displaystyle\left(\frac{\gamma-1}{\gamma}\frac{4\pi^2\kappa_\parallel}{\rho_0 C_\mathrm{V}}\right)k^2 + \frac{\tau_1-\tau_2}{\tau_1\tau_2}}.
\end{equation}

\subsection{Limit of strong non-adiabaticity}
\label{subsec:strong_limit}
The limit of strong non-adiabaticity describes slow magnetoacoustic waves for which $\omega \ll 1/\tau_{1,2,\mathrm{cond}}$, which is to say these waves are highly affected by the exchange of energy with the active medium. 
The dispersion relation (Eq.~\ref{eqn:dispersion_relation_final}) may be expressed in the following way (where we have divided through by $\omega \neq 0$):
\begin{equation}
       \omega^2 = C_\mathrm{T}^2 k^2 \frac{\displaystyle 1 + i\left\{\frac{1}{\gamma}\frac{4\pi^2}{\omega \tau_\mathrm{cond}} + \frac{1}{\omega \tau_1} \right\} }{\displaystyle 1 + i\left\{\frac{2}{2+\gamma \beta}\left(\frac{(1 + \beta/2) 4\pi^2}{\omega \tau_\mathrm{cond}} + \frac{1}{\omega \tau_2} + \frac{\gamma \beta }{2}\frac{1}{\omega \tau_1} \right) \right\} } 
    \label{eqn:misbalance_dispersion_ready4strong} 
\end{equation}
After Taylor expansion we find the strong limit to be 
\begin{equation}
 \begin{split} 
    \omega^2 \approx & \, C_\mathrm{S}^2 k^2 \left\{ \frac{\displaystyle \left(\frac{1}{\gamma}\frac{4\pi^2}{\tau_\mathrm{cond}} + \frac{1}{\tau_1}\right)}{\displaystyle \frac{4\pi^2}{\tau_\mathrm{cond}} + \frac{1}{\tau_2} +  \frac{\gamma\beta}{2}\left(\frac{1}{\gamma}\frac{4\pi^2}{\tau_\mathrm{cond}} + \frac{1}{\tau_1}\right)} \right. \\
    &\left. -i \omega  \frac{\displaystyle \left(\frac{\gamma -1}{\gamma}\right)\frac{4\pi^2}{\tau_\mathrm{cond}} + \frac{1}{\tau_2}-\frac{1}{\tau_1}}{\displaystyle \left(\frac{4\pi^2}{\tau_\mathrm{cond}} + \frac{1}{\tau_2} +  \frac{\gamma\beta}{2}\left(\frac{1}{\gamma}\frac{4\pi^2}{ \tau_\mathrm{cond}} + \frac{1}{\tau_1}\right)\right)^2 } \right\} .
 \end{split}
 \label{eqn:misbalance_strong}
\end{equation}
\noindent The change from $C_\mathrm{T}^2$ to $C_\mathrm{S}^2$ is caused by pulling out a factor of $1+\gamma \beta/2$.
Equation~(\ref{eqn:misbalance_strong}) agrees with strong limit in the infinite field limit as $\beta \rightarrow 0$ as it should \citep{Zavershinskii2019}.
In order to deal with the $\omega$ in the imaginary component, we again apply the perturbation approach. 
In the zeroth order $\omega$ is approximated by $\omega_\mathrm{R}$ as seen in Equation~(\ref{eqn:misbalance_strong}), yielding the following solution to the highly non-adiabatic dispersion relation:
\begin{align}
    &\omega_\mathrm{R} \approx \frac{1}{\sqrt{\gamma}} C_\mathrm{S} k \frac{\displaystyle \left(\frac{4\pi^2}{\tau_\mathrm{cond}} + \frac{\gamma}{\tau_1}\right)^{1/2}}{\displaystyle \left[\frac{4\pi^2}{\tau_\mathrm{cond}} + \frac{1}{\tau_2} + \frac{\beta}{2}\left(\frac{4\pi^2}{\tau_\mathrm{cond}} + \frac{\gamma}{\tau_1}\right)\right]^{1/2}} \ ,\label{eqn:misbalance_strong_Re} \\[2ex]
    &\omega_\mathrm{I} \approx -\frac{1}{2}C_\mathrm{S}^2 k^2 \left(1+ \frac{\gamma\beta}{2}\right) \frac{\displaystyle \left(\frac{\gamma-1}{\gamma}\frac{4\pi^2}{\tau_\mathrm{cond}}+\frac{1}{\tau_2}-\frac{1}{\tau_1}\right)}{\displaystyle  \left[\frac{4\pi^2}{\tau_\mathrm{cond}} + \frac{1}{\tau_2} +  \frac{\beta}{2}\left(\frac{4\pi^2}{\tau_\mathrm{cond}} + \frac{\gamma}{\tau_1}\right)\right]^{2}} \ .
    \label{eqn:misbalance_strong_Im}
\end{align}
We must make it clear that different combinations of signs of $\tau_1,\tau_2$ may lead to very different behaviour, e.g. complex phase speeds (that is, even non-propagating modes) or the development of thermal instabilities of a non-acoustic nature \citep[see][]{Field1965}. 
Restricting our attention to the case of a stable propagating slow wave ($\tau_1,\tau_2 >0$), in the absence of thermal conduction the non-adiabatic wave propagates at the speed $\omega_\mathrm{R}/k$ with $\omega_\mathrm{R} = C_\mathrm{S} k \left(\tau_2/ (\tau_1 + (\gamma \beta/2)\tau_2)\right)^{1/2}$, which in the infinite field case is simply $C_\mathrm{S}\sqrt{\tau_2/\tau_1}$. 
\emph{In non-zero $\beta$ plasma, the phase speed of the highly non-adiabatic wave is reduced compared to the infinite magnetic field case.}
The wave will damp if Equation~(\ref{eqn:misbalance_strong_Im}) is negative. 
The effect of non-zero $\beta$ upon this wave increment/decrement (that is, growth or damping) is governed by Equation~(\ref{eqn:misbalance_strong_Im}), and is different for different plasma conditions (equivalently its impact depends on the relative magnitudes of $\tau_\mathrm{cond}, \tau_1, \tau_2$). 
Unlike the weakly non-adiabatic case, the effects of the different non-adiabatic mechanisms upon the wave increment/decrement $\omega_\mathrm{I}$ are not additive. 

Considering only the thermal conduction terms ($\tau_{1,2} \gg \tau_\mathrm{cond}$), it is found that $\omega_\mathrm{R}/k \approx C_\mathrm{S} \left( \gamma(1 +  \beta/2)\right)^{-1/2}$. 
This tends to the isothermal sound speed $C_\mathrm{S}/\sqrt{\gamma}$ as $\beta \rightarrow 0$ consistent with \citet{DeMoortelHood2003_pt1}; as with the case for no thermal conduction, for increasing $\beta$ this phase speed is reduced. 
The effect of thermal conduction on the wave decrement is always to damp ($\omega_\mathrm{I}<0$).
However, as the wave approaches the isothermal regime (without misbalance) $\omega_\mathrm{I}$ becomes proportional to $\omega \tau_\mathrm{cond}$ which is a small parameter in the strong limit. 
In other words, although the effect of thermal conduction is always to damp, in the isothermal regime the damping ceases.
The effect of increasing $\beta$ is to reduce $\omega_\mathrm{I}$ and hence increase damping times (equivalent to lessening the rate of damping). 

\section{Damping of slow waves in the corona}
\label{sec:slow_mode_damping}

\subsection{Estimation of damping time for non-zero plasma-beta}
\label{subsec:estimate_damping_time}
We now focus on the damping effect of the thermal misbalance upon slow magnetoacoustic waves in the corona.
The characteristic timescales of wave-induced thermal misbalance (Eq.~\ref{eqn:misbalance_timescales}) vary with $T_0$, $\rho_0$, $B_0$, as well as the parameters dictating the heating and cooling rates $\mathcal{H}$ and $\mathcal{L}$.  
Up to this point, all our results have been expressed in terms of a generic heating/cooling function $Q$, whose derivatives with respect to thermal equilibrium are treated as free parameters, and applicable to any plasma conditions for which the governing equations may be satisfied. 
In order to fully explore our results in the coronal context however, we now pin down a functional form of $Q$ and pick some plasma parameter ranges to evaluate. 
We consider temperatures ranging from \SIrange{0.5}{20}{MK} and electron number densities ranging from \SIrange{1e8}{5e12}{\per\cm\cubed}, since many typical coronal structures have been detected at these temperatures and densities, such as plumes and coronal loops \citep{DeMoortel2009}. 
Some coronal structures do exist outside of these ranges, such as prominences, however for such conditions the effects of partial ionisation, non-LTE conditions and optical thickness can not be neglected. 

We parameterise the coronal heating/cooling function as 
\begin{equation}
    Q = \mathcal{L}(\rho,T) - \mathcal{H}(\rho,T,B) \Longleftarrow \left\{ \begin{aligned} &\ \mathcal{L}(\rho,T) \ \ \text{from CHIANTI}, \\
    &\ \mathcal{H}(\rho,T,B) = h_0 \rho^a T^b B^c.
    \end{aligned} \right.
    \label{eqn:misbalance_define_H_L}
\end{equation}
where the coefficient $h_0$ is determined from the initial thermal equilibrium condition, $Q_0=0$, $\implies h_0 = \mathcal{L}_0/\rho_0^a T_0^b B_0^c$, and the power indices $a$, $b$ and $c$ are treated as free parameters. 
We synthesise the coronal optically thin radiation function $\mathcal{L}(\rho,T)$ from CHIANTI atomic database v. 9.0.1 \citep{Dere1997,Dere2019} for the densities and temperatures from those intervals. 

We do not know the values of $a,b,c$, as this is essentially the coronal heating problem. 
As discussed in \citet{Kolotkov2020_constrainH}, many previous authors such as \citet{Ibanez1993,Carbonell2006} have considered five models for $a,b$ originating from appendix B of \citet[][]{Rosner1978}. 
However all five of these are incompatible with the observations of widespread coronal thermal stability and the rapid damping of slow (acoustic) waves. 
Following \citet{Kolotkov2020_constrainH} we consider the values of $a=1/2$, $b=-7/2$, for which both thermal stability and acoustic stability are always satisfied in coronal conditions, in accordance with observations. 
We study the change in damping with the parameter $c$, and the change in damping with plasma-$\beta$.

\begin{figure*} 
    \begin{subfigure}{0.5\textwidth}
        \includegraphics[width=\linewidth]{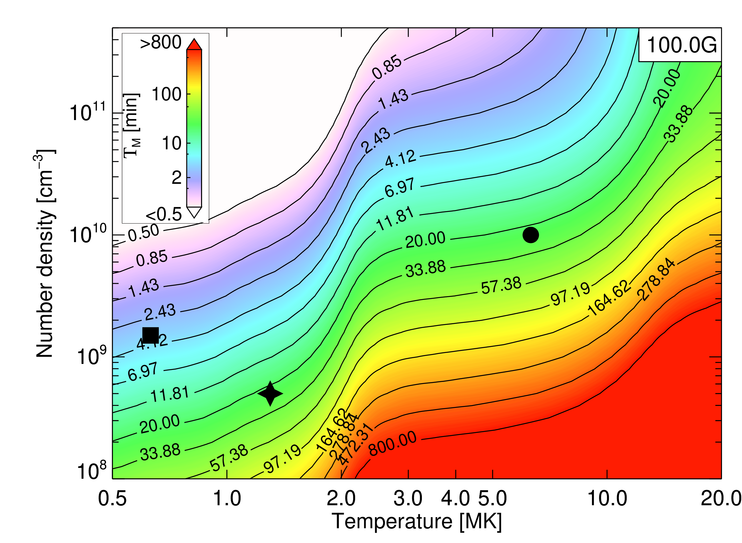}
        \caption{Plot of $\mathcal{T}_\mathrm{M}$ with $c=0$ at 100~G.} \label{fig:misbalance_c0B5000}
    \end{subfigure}\hspace*{\fill}
    \begin{subfigure}{0.5\textwidth}
        \includegraphics[width=\linewidth]{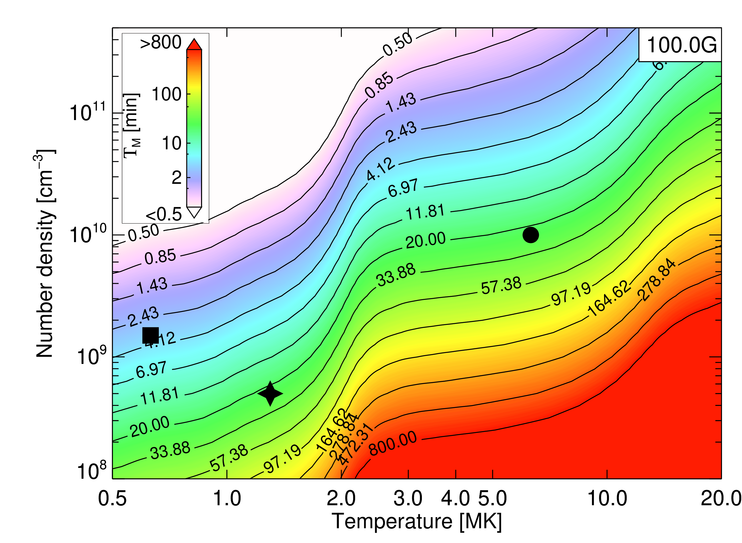}
        \caption{Plot of $\mathcal{T}_\mathrm{M}$ with $c=1$ at 100~G.} \label{fig:misbalance_c1B5000}
    \end{subfigure}
    
    \medskip
    \begin{subfigure}{0.5\textwidth}
        \includegraphics[width=\linewidth]{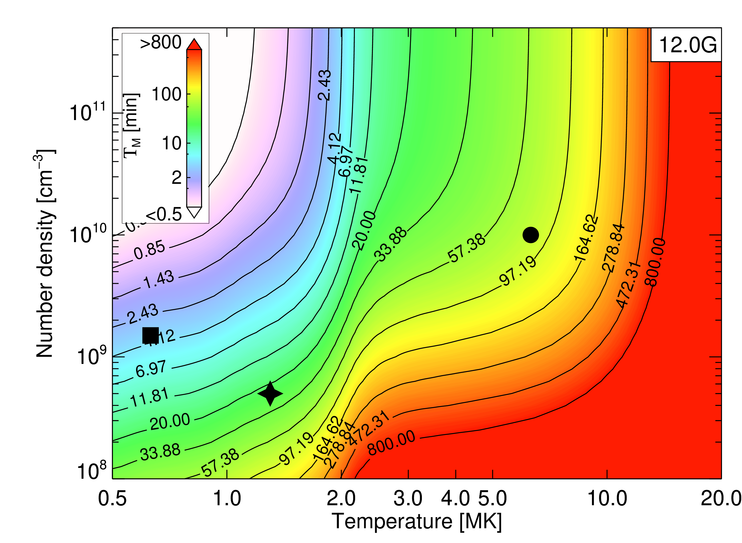}
        \caption{Plot of $\mathcal{T}_\mathrm{M}$ with $c=0$ at 12~G.} \label{fig:misbalance_c0B12}
    \end{subfigure}\hspace*{\fill}
    \begin{subfigure}{0.5\textwidth}
        \includegraphics[width=\linewidth]{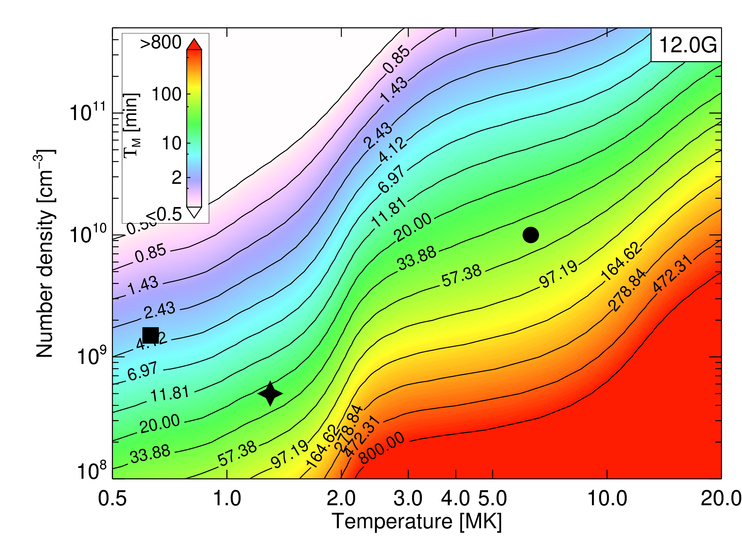}
        \caption{Plot of $\mathcal{T}_\mathrm{M}$ with $c=1$ at 12~G.} \label{fig:misbalance_c1B12}
    \end{subfigure}
    
    \medskip
    \begin{subfigure}{0.5\textwidth}
        \includegraphics[width=\linewidth]{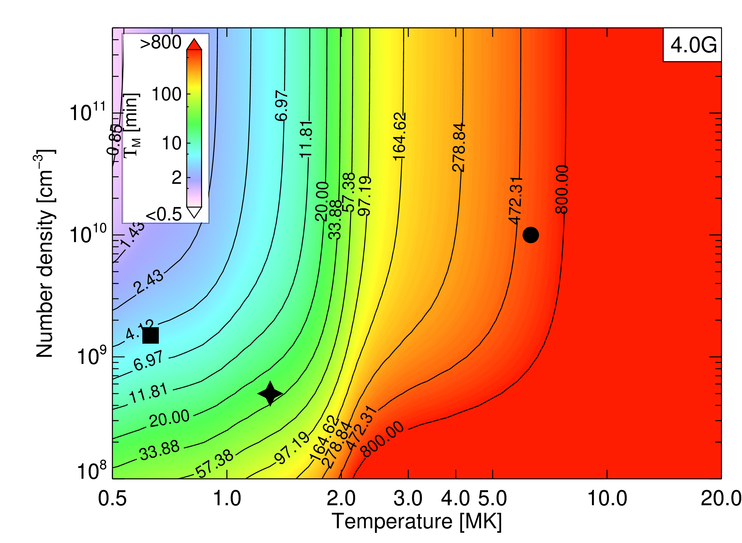}
        \caption{Plot of $\mathcal{T}_\mathrm{M}$ with $c=0$ at 4~G.} \label{fig:misbalance_c0B4}
    \end{subfigure}\hspace*{\fill}
    \begin{subfigure}{0.5\textwidth}
        \includegraphics[width=\linewidth]{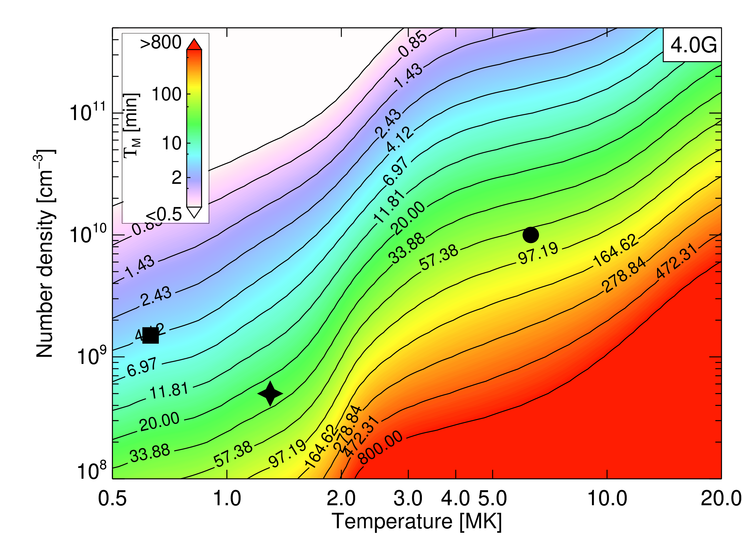}
        \caption{Plot of $\mathcal{T}_\mathrm{M}$ with $c=1$ at 4~G.} \label{fig:misbalance_c1B4}
    \end{subfigure}
    \caption{Variation of the characteristic thermal misbalance damping timescale $\mathcal{T}_\mathrm{M}$ with magnetic field, and with different power-law index $c$ where $\mathcal{H} \propto \rho^{1/2} T^{-7/2} B^c$. 
    Scanning down the column shows how the damping time changes as magnetic field $B_0$ decreases: panels (a), (b) at 100~G; panels {(c), (d)} at 12~G; panels (e), (f) at 4~G. 
    Comparing left-to-right shows the effect of a different power-law index $c$ (left side $c=0$, right side $c=1$), whilst all other parameters are held the same. 
    Symbols correspond to specific plasma conditions (see Table~\ref{tab:misbalance_timescales}).
    Note that panels (a) and (b) are practically identical, since the plasma-$\beta$ everywhere in these plots is sufficiently close to zero for the infinite magnetic field approximation to apply, which is independent of $\partial Q/\partial B$ (see Subsec.~\ref{subsec:inf_limit}).
    } \label{fig:misbalance_c0c1_effects}
\end{figure*}

We now estimate the absolute values of the characteristic thermal misbalance damping time $\mathcal{T}_\mathrm{M}$ in the limit of weak non-adiabaticity (recall the connection to the wave damping time $\mathcal{T}_\mathrm{damp} = 2\mathcal{T}_\mathrm{M}$, in the absence of damping by thermal conduction).
The effect of a weaker magnetic field (i.e. higher plasma-$\beta$) on $\mathcal{T}_\mathrm{M}$ may be seen in Figure~\ref{fig:misbalance_c0c1_effects}, for two heating models with different magnetic dependencies (chosen for illustration purposes only). 
Looking only at the effect of reduced magnetic field (scanning downwards in Fig.~\ref{fig:misbalance_c0c1_effects}) across the range of typical coronal magnetic field strengths, the damping time decreases due to non-zero $\beta$ effects, particularly temperatures over $\sim 2$~MK; the change in magnetic field strength has the most pronounced effect on hotter, denser plasma (i.e. where plasma-$\beta$ is greater) whereas the cooler loops and plumes remain largely unaffected. 

There is a distinction between the impact of the finite magnetic field (non-zero plasma-$\beta$) upon the wave-induced thermal misbalance, and the effect of any dependence of the heating/cooling function upon magnetic field strength (non-zero $Q_B$). 
To demonstrate the latter of these, consider the difference introduced by the change in dependence of $\mathcal{H}$ upon $B$ (scanning left to right in Fig.~\ref{fig:misbalance_c0c1_effects}). 
For these heating functions the magnetic heating power-law index ($c=0 \rightarrow c=1$) has made the damping time vary \textit{less} with magnetic field strength.
In other words, the heating scenario with $c=1$ has a stabilising effect on the wave dynamics. 
This may not be the case for other values of $c$ (e.g. $c=-1$), this is not discussed in this work.
The difference is apparent only for lower magnetic field strengths. 
At infinite magnetic field, there is no difference between damping times for the two heating models -- see panels (a) and (b) in Figure~\ref{fig:misbalance_c0c1_effects}, which are almost identical.
This implies that \emph{above a certain magnetic field strength, the infinite magnetic field approximation is valid regardless of the heating functional dependence upon magnetic field.}

\subsection{Sensitivity of the wave damping to the dependence of heating function upon magnetic field}
\label{subsec:inf_limit}
\begin{figure}[ht!]
	\begin{center}
	    \resizebox{\hsize}{!}{\includegraphics{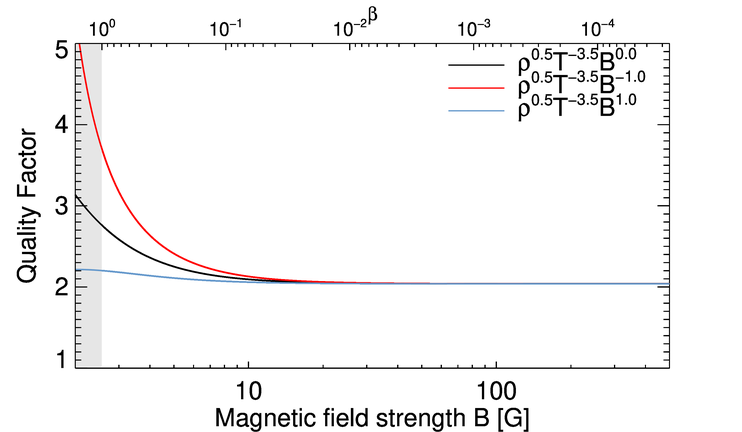}} \\[2ex]
    	\resizebox{\hsize}{!}{\includegraphics{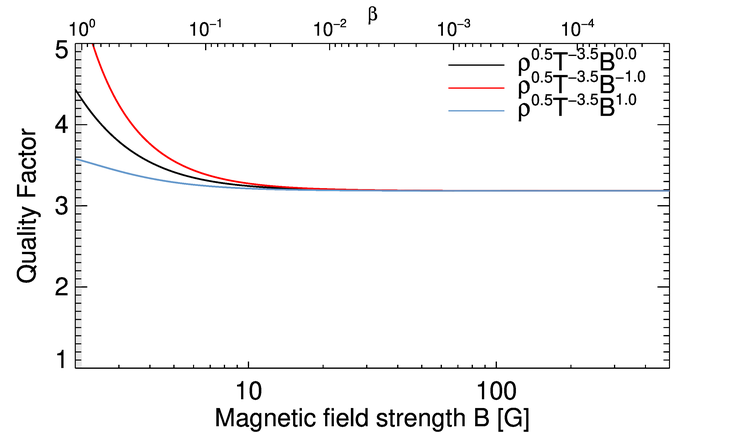}} \\[2ex]
    	\resizebox{\hsize}{!}{\includegraphics{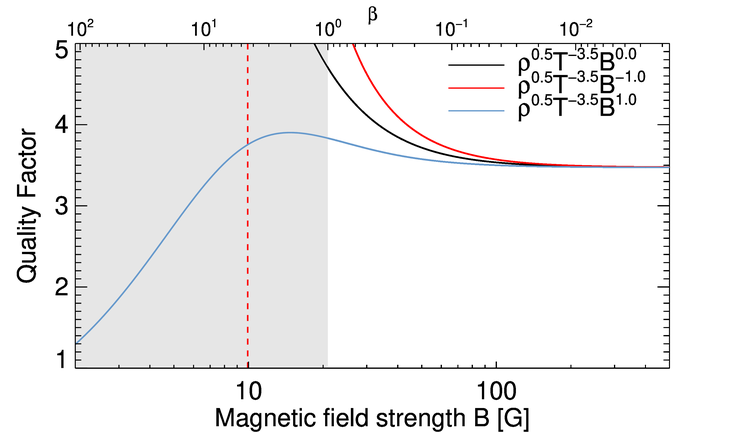}}
	\end{center}
	\caption{Plots of damping time over period (quality factor) against magnetic field strength, for 3 heating models with differing dependencies on magnetic field strength, which are each coloured (Eqs.~\ref{eqn:weak_damping_time}, \ref{eqn:misbalance_weak_Re}). 
	[\textit{Top}] Quality factors calculated for plasma parameters corresponding to an upwardly propagating slow wave in a coronal fan loop above a sunspot, $T$ = \SI{0.63}{MK}, electron density of $n_e$ = \SI{1.5e9}{cm^{-3}}, and periodicity set to 3.0~minutes corresponding to a wavelength of $\lambda$ = \SI{22}{Mm}. 
	[\textit{Middle}] Quality factors calculated for a slow wave propagating in a coronal plume, $T$ = \SI{1.3}{MK}, $n_e$ = \SI{5e8}{cm^{-3}}, and wavelength $\lambda$ set to \SI{100}{Mm} corresponding to a periodicity of 9~minutes. 
	[\textit{Bottom}] Quality factors calculated for a standing slow wave in a hot loop observed by SUMER, $T$ = \SI{6.3}{MK}, $n_e$ = \SI{1e10}{cm^{-3}}, $\lambda$ = \SI{250}{Mm} yielding a periodicity of 11~minutes. 
	Grey shading marks the $\beta >1$ region, and the dotted line in the bottom plot marks a region in which $\mathcal{T}_\mathrm{M} <0$ and so the effect of thermal misbalance is destabilising the plasma. 
	}
	\label{fig:qf_vs_B}
\end{figure}
In order to estimate the magnetic field strength above which the infinite magnetic field approximation is appropriate, we consider the combined damping effect of thermal misbalance and thermal conduction. 
This is necessary because the thermal conduction damping term is also affected by non-zero plasma-$\beta$ (see Eq.~\ref{eqn:tau_condB}), so its exclusion would not allow for delineating a complete picture.
For simplicity, we only consider the weakly non-adiabatic limit, such that the angular frequency of the slow wave is approximated by Equations~(\ref{eqn:misbalance_weak_Re}), (\ref{eqn:misbalance_weak_Im}) and (\ref{eqn:tau_M_adds}). 
We specify three pertinent examples of damped slow waves seen in the corona: the 3-minute upwardly propagating slow waves above a sunspot observed in the SDO/AIA \SI{171}{\angstrom} bandpass which peaks at \SI{0.63}{MK} \citep{DeMoortel2009}, the slow waves seen in a plume observed with the \SI{193}{\angstrom} bandpass (which peaks at \SI{1.3}{MK}) \citep[e.g.][]{Prasad2014}, and the standing oscillations in hot loops observed by SUMER, with a formation temperature of Fe XIX of \SI{6.3}{MK}) \citep[e.g. recently reviewed in][]{Kosak2019}. 
As before we use the illustrative choice of heating scenario taken from \citet{Kolotkov2020_constrainH}, $\mathcal{H} \propto \rho^{1/2} T^{-7/2} B^c$, and let $c$ vary from -1, 0, 1. 

Figure~\ref{fig:qf_vs_B} shows that for sufficiently strong magnetic field strengths, slow waves in the presence of all of the heating models shown have converged to the quality factor calculated for the infinite magnetic field case.
This is regardless of the dependence of heating model upon magnetic field, in our case controlled by the power-law index $c$.
For the warm quiescent corona, as demonstrated by the top two panels, a magnetic field strength greater than around \SI{10}{G} is sufficient for the infinite magnetic field approximation to be appropriate. 
For the conditions typical of hot loops seen by SUMER, which are often post-flare, the non-zero $\beta$ effects are still important at high magnetic field strengths, and so a magnetic field strength of approximately one order of magnitude higher ($\sim$\SI{100}{G}) is required for the damping to be independent of the heating/cooling function's dependence upon $B$. 
This difference in behaviour may also be seen in Figure~\ref{fig:misbalance_c0c1_effects} since it is the hot, dense plasmas (upper right quadrants of those plots) for which the change with plasma-$\beta$ and with the change in magnetic dependence $c$ is greatest and visible.
Returning to Figure~\ref{fig:qf_vs_B}, it may also be seen that if $Q_B = 0$ (black lines) or the plasma-$\beta$ is sufficiently small to neglect the effects of $Q_B$, then the damping is always diminished with increasing $\beta$ (stronger magnetic fields mean more damping) -- consistent with Subsection~\ref{subsec:weak_limit}.

When the plasma-$\beta$ greatly exceeds 1, that is to say when the plasma is pressure dominated as opposed to magnetically dominated, the timescales for both the thermal misbalance and the thermal conduction depart greatly from the infinite magnetic field values. 
This is apparent in the bottom panel of Figure~\ref{fig:qf_vs_B}, where the quality factors have diverged greatly in the $\beta>1$ region. 
Also, the effect of plasma-$\beta$ upon the wave damping can now be to decrease or increase the damping, depending on the sign of $Q_B$. 
We do not consider the regime $\beta>1$ since it is more applicable to chromospheric plasma and below, necessitating the addition of further physical effect such as optically thick radiation, partial ionisation and non-LTE conditions.

\subsection{Comparison with damping by thermal conduction}

\begin{table*}[t]
\centering
    \begin{tabular}{|C{4cm}|| C{2.9cm} C{2.9cm} C{2.9cm}|} 
     \hline
     Typical value & Loop in \SI{171}{\angstrom} & Plume in \SI{193}{\angstrom} & Hot loop in SUMER \\ [0.5ex]
     (Symbol on plots) & (square) & (star) & (circle) \\ [0.5ex]
     \hline\hline
    Temperature, $T_0$ & 0.63~MK & 1.3~MK & 6.3~MK \\[0.2ex] 
	Number density, $n_e$ & \SI{1.5e9}{cm^{-3}} & \SI{0.5e9}{cm^{-3}} & \SI{1e10}{cm^{-3}} \\ [0.2ex]
	Period, $P$ &  $\approx 3-10$~min & $\approx 8-18$~min & $\approx 10-20$~min\\ 
	\hline 
	$\mathcal{T}_\mathrm{M}$, $B=\infty$ & 3.2~min & 23~min & 23~min \\[1ex] 
    $\mathcal{T}_\mathrm{M}$, $B=34~G$ & 3.2~min \newline {\small($\beta$=0.01)} & 23~min \newline {\small($\beta$=0.00)} & 30~min \newline {\small($\beta$=0.38)}  \\[0.2ex] %
    $\mathcal{T}_\mathrm{M}$, $B=12~G$ & 3.3~min \newline {\small($\beta$=0.05)} & 24~min \newline {\small($\beta$=0.03)} & 81~min \newline {\small($\beta$=3.0)} \\ [0.2ex]%
    $\mathcal{T}_\mathrm{M}$, $B=4~G$ & 4.4~min \newline {\small($\beta$=0.43)} & 29~min \newline {\small($\beta$=0.28)} & 548~min $\simeq$ 9~hr \newline {\small($\beta$=27)} \\[0.2ex] %
    \hline
 	$\mathcal{T}_\mathrm{condB}$, $B=\infty$ & 10~min & 11~min & 27~min \\ [1ex] 
	$\mathcal{T}_\mathrm{condB}$, $B=34~G$ & 10~min \newline {\small($\beta$=0.01)} & 11~min \newline {\small($\beta$=0.00)} & 36~min  \newline {\small($\beta$=0.38)} \\ [0.2ex]%
    $\mathcal{T}_\mathrm{condB}$, $B=12~G$ & 11~min \newline {\small($\beta$=0.03)} & 12~min \newline {\small($\beta$=0.03)}  & 97~min  \newline {\small($\beta$=3.0)} \\ [0.2ex]%
    $\mathcal{T}_\mathrm{condB}$, $B=4~G$ & 14~min \newline {\small($\beta$=0.43)} & 14~min \newline {\small($\beta$=0.28)} & 650~min $\simeq$ 11~hr  \newline {\small($\beta$=27)} \\ [0.2ex]%
     \hline
     \hline
    $\tau_\mathrm{rad}$ & 9.7~min & 60~min & 70~min \\ [0.2ex]
    \hline
    \end{tabular}
    \caption{Table comparing the characteristic timescales calculated for the typical values of three coronal plasma non-uniformities in which rapidly decaying slow modes have been observed: 
    warm quiescent loops seen in \SI{171}{\angstrom} \citep[][]{DeMoortel2009}, 
    coronal plumes seen in \SI{193}{\angstrom} \citep[e.g.][]{Prasad2014}, 
    and hot dense loops observed in the Fe XIX channel by SUMER \citep{Kosak2019}.  
    The three points ($T,\rho$) are marked on the plots in Fig.~\ref{fig:misbalance_c0c1_effects}, and are the same parameters used for the three plots in Fig.~\ref{fig:qf_vs_B}.  
    The characteristic timescale $\mathcal{T}_\mathrm{M}$ is calculated using Eq.~(\ref{eqn:tau_M}) for heating model $\mathcal{H} = \rho^{1/2} T^{-7/2}$.
    The thermal conduction damping time $\mathcal{T}_\mathrm{condB}$ is calculated from Eq.~(\ref{eqn:tau_condB}) using wavelengths $\lambda$ = \SIlist{22;100;250}{Mm} respectively.
    A range of magnetic field strengths (hence $\beta$) are presented.
    The characteristic radiative timescale $\tau_\mathrm{rad}$ is calculated from Eq.~(\ref{eqn:misbalance_tau_rad})
    \label{tab:misbalance_timescales}
    }
\end{table*}
As Figure~\ref{fig:misbalance_c0c1_effects} demonstrates, the damping timescale for the thermal misbalance in the infinite magnetic field case is of the same order as the observed periodicity of slow waves in many typical coronal conditions, and the same order again as the observed damping times (some tens of minutes). 
To check this holds true when accounting for non-zero $\beta$ plasma, as well as compare the damping by thermal misbalance with the (conventionally dominant) damping by thermal conduction, we calculate the characteristic damping times due to thermal misbalance $\mathcal{T}_\mathrm{M}$ and $\mathcal{T}_\mathrm{condB}$ for typical combinations of coronal densities, temperatures and magnetic field strengths. 
The results are shown in Table~\ref{tab:misbalance_timescales}. 

We stress that the variation of thermal conduction damping time $\mathcal{T}_\mathrm{condB}$ with both $\beta$ and $\lambda$ means that its relevance is extremely broad, and the thermal misbalance timescale $\mathcal{T}_\mathrm{M}$ depends on the exact parameterisation of the as-yet unknown coronal heating function. 
Even so, for typical coronal situations Table~\ref{tab:misbalance_timescales} leads us to conclude that \emph{when comparing the damping of slow waves by thermal misbalance with the damping by field-aligned thermal conduction, we find the effect of the heating/cooling misbalance could be of equal or greater importance}. 
As a specific example, consider the propagating 3~minute oscillations seen in \SI{171}{\angstrom} with a wavelength $\lambda \approx 22$~Mm. 
Both $\mathcal{T}_\mathrm{M}$ and $\mathcal{T}_\mathrm{condB}$ are of the same order as the wave period, and the quality factor calculated for this combination of parameters is $\sim$2 (top panel of Fig.~\ref{fig:qf_vs_B}). 
This is consistent with the observations of the rapid damping of these propagating slow waves. 

The evaluation of the effect of the heating and cooling upon the slow wave may easily be confused with the cooling timescale of the host plasma, often defined as \citep[see e.g. Eq. (6) in][]{DeMoortelHood2004_pt2}
\begin{align}
    &\tau_\mathrm{rad} = \frac{\gamma C_\mathrm{V}T_0}{\mathcal{L}_0(\rho_0, T_0)}. \label{eqn:misbalance_tau_rad}
\end{align}
Although the values of timescales $\tau_\mathrm{rad}$ in Table~\ref{tab:misbalance_timescales} look similar to their counterparts the misbalance damping timescales $\mathcal{T}_\mathrm{M}$, from a physical point of view they are independent and describe fundamentally different processes. 
The quantity $\tau_\mathrm{rad}$ is associated with the host plasma (not the wave), it depends on the magnitude of the radiative losses, and neglects the influence of coronal heating which indisputably exists. 
The cooling with the characteristic time $\tau_\mathrm{rad}$ occurs when the heating of the plasma is suddenly switched off.
In contrast, the characteristic timescales $\tau_1,\tau_2$ and damping time $\mathcal{T}_\mathrm{M}$ are determined by the \emph{derivatives} of the complete heating/cooling function and the plasma parameters, and characterise the effect of the wave-induced misbalance upon the wave when both cooling and heating processes are still operating. 
Thus, the radiative timescale does not reflect if the effect of misbalance between heating and cooling is important for the slow magnetoacoustic wave: the heating/cooling misbalance may have a great effect even if $\tau_\mathrm{rad}$ is far from the wave period $\omega$. 

The slow waves considered in Table~\ref{tab:misbalance_timescales} all lie comfortably in the weakly non-adiabatic regime, $\omega\times\{ \tau_\mathrm{cond},\tau_1,\tau_2 \} \gg 1$. 
This may not necessarily be true for all slow waves in the corona.
Supposing the thermal conduction were strong enough to be in the strongly non-adiabatic regime 
$\omega \tau_\mathrm{cond} \ll 1$, the damping time should be calculated using Equation~(\ref{eqn:misbalance_strong_Im}) and tends to no damping by thermal conduction in the isothermal limit \citep{DeMoortelHood2003_pt1}.
In contrast, thermal misbalance may cause strong damping even in the isothermal regime in which the conductive damping is very weak, via the wave's perturbations to density and magnetic field, not temperature.
This makes the damping by thermal misbalance a viable mechanism for damping in isothermal regimes.

\section{Discussion and Conclusions}
\label{sec:discussion}

The importance of non-adiabatic effects for the damping of slow modes has been shown in many previous works, however in some cases the importance of the presence of steadily operating coronal heating for slow modes has not been realised because the heating term is considered a constant (that is, \textit{unperturbed}) \citep[e.g.][]{DeMoortelHood2004_pt2}. 
As we have shown, if the coronal heating mechanism is acting during the oscillation, then the damping effect of wave-induced misbalance between the heating and cooling mechanisms can be significant and should not be neglected. 
We must stress that in our study the energy for heating does not come from the slow wave, and is supplied by some other mechanism. 

The potential for the inclusion of damping by thermal misbalance to explain the various discrepancies between observed slow mode damping and theory 
is clearly enormous. 
In the weakly non-adiabatic limit, the damping (or amplification) by heating/cooling misbalance does not change with wavenumber $k$, meaning its effect is universal for different length structures. 
In the general non-adiabatic case, there is some dependence upon $k$. 
The damping by thermal conduction always varies with length scale. 
Thus if the slow wave is damped by both thermal conduction and thermal misbalance (e.g. Eq.~\ref{eqn:weak_damping_time}), the dependence of damping time upon frequency would be more complicated than a straight line of gradient 2 on a log-log plot (as was previously expected since $\tau_\mathrm{cond} \propto k^{-2}$). 
The inclusion of thermal misbalance as a damping mechanism may therefore explain the unexpected frequency dependencies found in \citet{Prasad2014}, since the gradients of best fit on period vs damping length plots can take a range of values depending on the relative contributions of $\mathcal{T}_\mathrm{M}$ and $\mathcal{T}_\mathrm{condB}$. 
Moreover, the difference seen between the damping of slow waves observed in plumes and those seen in sunspots \citep{Prasad2014}; the unexpected dependence of damping length upon temperature reported in \citet{Prasad2019}; and the change of the damping's frequency dependence with height in \citet{Gupta2014}
are naturally explained by the variation of $\tau_\mathrm{damp}$ upon plasma-$\beta$ and/or the variation of $\mathcal{T}_\mathrm{M}$ with density, temperature and heating function. 
Furthermore, the thermal misbalance will likely introduce a phase shift between density and temperature in the same manner as thermal conduction does \citep[see Sec 3.1.2,][]{Owen2009} which may explain the results in e.g. \citet{Prasad2018}. 
The variation of this phase shift with the effectiveness of the thermal misbalance would have the consequence of making measurements of the coronal polytropic index, measured using density/temperature phase shifts, also vary with the effectiveness of the thermal misbalance. 
As we have demonstrated in this work \citep[see also][]{Kolotkov2020_constrainH} the effect of thermal misbalance varies with temperature in the corona, and so one may expect that the polytropic index would vary with temperature as was the case in \citet{Prasad2019}, though further validation is warranted.

The damping and dispersion of slow waves in the corona by wave-induced thermal misbalance are subject to non-zero $\beta$ effects, some of which are irrespective of the heating/cooling function whilst further effects may occur if there is any dependence of $Q$ upon magnetic field.
Regarding purely non-zero $\beta$ effects, we have found that \emph{a wave propagating through a non-zero $\beta$ plasma will always have a reduced phase speed compared to a wave in the infinite magnetic field case}, whilst its effect on the wave attenuation depends on the exact plasma conditions. 
In the case of a damped wave in weakly non-adiabatic plasma, any \emph{reduction in the magnetic field strength (increase in plasma-$\beta$) will increase the damping time}. 
The effect of thermal conduction is diminished (compared to the infinite field case) as the plasma-$\beta$ grows, such that both the isothermal phase speed and the damping rate from thermal conduction are reduced.
Regarding the effects of any dependence of $Q$ upon of magnetic field strength, one important effect on slow waves which may be important even for low-$\beta$ plasma, is the stability of the plasma to the isentropic instability (and potentially the thermal instability as well). 
In this work we have focussed on the damping effect of thermal misbalance upon slow waves, enforced by our choice of heating model informed by \citet{Kolotkov2020_constrainH}, since such heating model(s) is chosen such that slow modes are damped everywhere in the corona.
This may not be the case everywhere, since this assumption renders the phenomenon of coronal rain from thermal instability an impossibility, which is evidently not true. 
However the topic of instability in a non-zero $\beta$ plasma with magnetically dependent heating will be the subject of its own dedicated work in the future.

One major result of this work is that the infinite magnetic field approximation is good for the quiescent corona when the magnetic field strength is above $\sim$\SI{10}{G}. 
The magnetic field strengths in coronal structures are difficult to observe directly, often relying on seismological inference. 
Typical values in transversely oscillating coronal loops lie in the tens of Gauss \citep[e.g. see the inferences in tables B.1 and B.2 in][]{Arregui2019_inferB}, but values of several kilogauss have been reported at the base of the corona above exceptionally strong sunspots \citep[e.g.][]{Anfinogentov2019_bigB}.
Thus it may be concluded that, for the majority of the quiescent corona, the effects of the dependency of the heating model upon magnetic field strength may be safely neglected, and the infinite magnetic field approximation used instead.
The situation may be different in particularly hot loops such as after a flare, where the plasma-$\beta$ tends to be higher, and it is in these non-zero $\beta$ regions in which any dependence of the heating/cooling function upon magnetic field strength may be probed in a manner analogous to \citet{Kolotkov2020_constrainH}.
A natural generalisation of our study would be the consideration of the heating scenario which depend also upon the height above the bottom of the corona. 

The key results of this paper may be summarised into the following:
\begin{enumerate}
    \item The dispersion relation governing slow magnetoacoustic waves along an infinitely thin cylinder with non-zero $\beta$ was derived. 
    Crucially, two timescales ($\tau_1$ and $\tau_2$) that characterise the effect of wave-induced thermal misbalance are generalised for the non-zero $\beta$ case \citep[these timescales were found in the infinite magnetic field case in e.g.][]{Kolotkov2019}. 
    These are inversely proportional to the combined heating/cooling functions' derivatives with respect to temperature at constant \textit{gas} pressure, and with respect to temperature at constant \textit{magnetic} pressure respectively (Eq.~\ref{eqn:misbalance_timescales}). 
    \item The effect of heating/cooling misbalance in the limit of weak non-adiabaticity was found, applicable for waves in which the exchange of energy with the medium is only mild. 
    Such waves propagate at the tube speed $C_\mathrm{T}$, and their amplitude damping may be calculated through Equation~(\ref{eqn:tau_M_adds}). 
    In this limit the two characteristic timescales for thermal misbalance may be combined into a single damping time $\mathcal{T}_\mathrm{M}$, whose effect on the wave decrement is additive to that from thermal conduction (Eqs.~(\ref{eqn:misbalance_ready4weak})--(\ref{eqn:misbalance_weak_Im})). 
    The sign of $\mathcal{T}_\mathrm{M}$ may be positive (enhanced damping) or negative (reduced damping or over-stability).
    A change in magnetic field strength (plasma-$\beta$) will change the damping rate depending on the sign of $Q_B$. 
    If $Q_B =0$ (or the magnetic field is sufficiently strong for the infinite field approximation to be valid), then a decrease in magnetic field strength (increase in plasma-$\beta$) will always lessen the damping rate. 
    Thermal conduction always acts to damp the wave, and its effect is also reduced as $\beta$ increases. 
    \item The effect of thermal misbalance in the limit of strong non-adiabaticity was found, applicable for waves in which the exchange of energy with their medium is extreme (Eqs.~(\ref{eqn:misbalance_strong})--(\ref{eqn:misbalance_strong_Im})). 
    In this limit, the effects upon the wave decrement by parallel thermal conduction and by thermal misbalance are not additive. The limited isothermal phase speed in this regime is reduced for greater $\beta$.  
    \item The damping effect of wave-induced thermal misbalance upon slow magnetoacoustic waves is important for a wide range of coronal conditions, demonstrated through Table~\ref{tab:misbalance_timescales} by reason of the heating/cooling misbalance's damping timescale $\mathcal{T}_\mathrm{M}$ coinciding with typical observed coronal slow wave periods and damping times. 
    The damping by thermal misbalance is of comparable importance to the damping effect by thermal conduction.
    The different physical origins (and therefore different parametric dependencies) of these omnipresent damping mechanisms may explain the discrepancies reported between observations of slow mode damping in the corona and theory. 
    \item The quality factors for 3 minute slow mode oscillations above sunspots, slow modes in plumes, and in hot (post-flare) loops are estimated, considering both damping both thermal conduction and damping by wave-induced thermal misbalance. 
    For sufficiently large $\beta$ plasma, the damping of slow waves is independent of the heating/cooling functional dependence upon magnetic field. 
    As a rule of thumb, the infinite magnetic field approximation is valid for studying the effect of thermal misbalance in the quiescent corona for magnetic field strengths greater than $\approx$ \SI{10}{G}.
\end{enumerate}

\begin{acknowledgements}
The work was supported by the STFC consolidated grant ST/T000252/1. 
D.Y.K. acknowledges support from the budgetary funding of Basic Research program No. II.16.
V.M.N. acknowledges the Russian Foundation for Basic Research grant No. 18-29-21016.
CHIANTI is a collaborative project involving George Mason University, the University of Michigan (USA), and the University of Cambridge (UK).
\end{acknowledgements}
%
\bibliographystyle{aasjournal} 
\bibliography{bibliography}

\end{document}